\documentclass[apjl]{emulateapj}
\usepackage{color}

\definecolor{brown}{rgb}{0.42,0.24,0.07}

\newcommand{\mearth}{\,$M_{\oplus}$}
\newcommand{\mearthp}{\,$M_{\oplus}$ }

\shorttitle{Formation of planetary cores}
\shortauthors{S\'andor et al.}

\begin{document}


\title{Formation of planetary cores at Type I migration traps}

\author{Zsolt S\'andor\altaffilmark{1}, Wladimir Lyra\altaffilmark{2}, and Cornelis P. Dullemond\altaffilmark{1,3}}
    \email{sandor@mpia.de}



\altaffiltext{1}{Max Planck Research Group, Max-Planck-Institut f\"ur Astronomie, K\"onigstuhl 17, 69117, Heidelberg, Germany}
\altaffiltext{2}{Department of Astrophysics, American Museum of Natural History, 
    79th Street at Central Park West, New York, NY 10024, USA}
\altaffiltext{3}{Institut f\"ur Theoretische Astrophysik, Universit\"at Heidelberg, Heidelberg, Germany}

\begin{abstract}
  One of the longstanding unsolved problems of planet formation is
  how solid bodies of a few decimeters in size can ``stick'' to form large
  planetesimals. This is known as the ``meter size barrier''. In recent
  years it has become increasingly clear that some form of ``particle
  trapping'' must have played a role in overcoming the meter size barrier.
  Particles can be trapped in long-lived local pressure maxima, such as
  those in anticyclonic vortices, zonal flows or those believed to occur
  near ice lines or at dead zone boundaries. Such pressure traps are the
  ideal sites for the formation of planetesimals and small planetary
  embryos. Moreover, they likely produce large quantities of such bodies in
  a small region, making it likely that subsequent N-body evolution may lead
  to even larger planetary embryos. The goal of this Letter is to show that
  this indeed happens, and to study how efficient it is. In particular, we
  wish to find out if rocky/icy bodies as large as 
  10 \mearthp can form within 1 Myr, since such bodies are the 
  precursors of gas giant planets in the core accretion scenario.
\end{abstract}


\section{Introduction}

There are a number of major unsolved problems in the theory of rocky planet
formation. One is the infamous ``meter size barrier'' problem
\citep{blum&wurm08}: When dust aggregates reach sizes beyond roughly 1 cm,
they partially decouple dynamically from the gas and can reach relative
velocities with respect to other particles of up to 50 m/s
\citep{weidenschilling&cuzzi1993}.  
Any such collision will likely lead
to fragmentation and/or erosion rather than growth. In addition, these
bodies tend to drift radially inward toward the star at a high speed, and
thus quickly get lost to the planet formation process
\citep{braueretal08}. Moreover, as laboratory experiments and numerical
modeling \citep{zsometal10} show, already at millimeter sizes the dust
aggregates stick insufficiently well for coagulation to continue. Despite
many theoretical studies aimed at solving this problem, this issue is still
wide open. In recent years a new line of thought has emerged which invokes
various local particle trapping mechanisms to solve it. \citet{cuzzietal08}
propose that particle concentrations in a turbulent disk occurring naturally
at small eddy scales can, statistically, sometimes lead to self-gravitating
``sandpiles'' that gradually condense to form planetesimals of 1 to 100 km
radius. The particles must have grown to millimeter sizes before this
process sets in, but the meter size barrier is thus easily
overcome. \citet{johansenetal07} showed that particle trapping and
subsequent gravitational contraction of particle clouds can also happen at
the scale of the largest turbulent eddies, leading to bodies between 100 km
and 1000 km in size. This mechanism can in fact also act on scales much
larger than the turbulent eddy scale. For instance,
\citet{barge&sommeria1995} and \citet{klahr&henning1997} showed that
particles tend to get trapped in anticyclonic vortices, if they exist in
protoplanetary disks. \citet{varniere&tagger06} and \citet{terquem08} showed
with alpha disk models \citep{shakura&sunyaev73} that pressure enhancements
are expected at dead zone boundaries because of the difference in turbulent
viscosity. Such enhancements were confirmed in the MHD simulations of
\citet{katoetal2010}. \citet{kretke&lin07} suggested that similar pressure
ridges form at the snow line in the disk \citep[see discussion
by][]{dzyurkevichetal2010}, while \citet{johansenetal09} showed that an
``inverse cascade'' of magnetorotationally driven turbulence will lead to
large scale pressure bumps in disks.

Another extensively studied problem is the retention of protoplanets in the
disk for a sufficiently long time for the core accretion process. This
requires at least 10 times slower type I migration speed than estimated from
analytical theory \citep{alibertetal05}. In recent years, however, the
understanding of type I migration has changed considerably
\citep[e.g.][]{paardekoopermellema06,kleyetal09,paardekooperetal10}.  As
\citet{morbietal08} have demonstrated by using a proper surface density
profile in their hydro simulations a planet trap appears, which helps
forming massive rocky/icy bodies and prevents their migration to the central
star. We will hereafter adopt the (non-standard) nomenclature ``terrestrial
planet'' and ``planetary core'' for rocky/icy bodies of mass below and above
10 \mearth, respectively.

Many of the above ideas were combined in a single model by
\citet{lyraetal08,lyraetal09}. The model explores what happens at dead zone
boundaries, where the density enhancement was predicted by
\citet{varniere&tagger06} and \citet{inaba&barge06} to be unstable to the
Rossby Wave Instability \citep{lovelaceetal99}, which in turn leads to the
formation of large scale anticyclonic vortices. Particles of about cm to
meter in size subsequently drift into the vortices and form gravitationally
bound clumps of solids, ranging between the masses of the Moon and Mars.

The present paper follows up on that work. \citet{lyraetal08,lyraetal09}
explored only the formation mechanism, since the hydrodynamical models are
too burdensome to follow the evolution for longer than a few thousand of
years (which is but a small fraction of the lifetime of the gas-rich phase
of the disk). A large number of massive embryos are formed that all
initially have semi-major axes very close to the radial location of the
pressure bumps where they were formed. The location near these ``planet
factories'' are thus quickly overpopulated with embryos. Mutual collisions
and merging events are naturally expected in such scenario. It would be of a
great interest to determine how the ensemble develops over a long time
scale.

In this paper we follow the N-body evolution of the heaviest embryos of the swarm 
that were produced during the first few hundred orbits in the above model. We
include the gravitational interaction between the embryos and the gas in the
disk which leads to type I migration. We account for it by applying the 
analytic formulae used by \citet{lyraetal10}, and developed originally by 
\citet{paardekooperetal10}. The model is detailed in the next section. 

\section{Implementation of type I migration}

To integrate numerically the differential equations of the gravitational N-body 
problem we developed a Bulirsch-Stoer integrator, which can handle collisions 
between nearby bodies. When the mutual distance between the center of mass of two 
nearby bodies is less than the sum of their physical radii, the bodies 
collide and merge. The physical radius of a body is calculated using its mass 
assuming a $2\ \mathrm{g/cm^3}$ bulk density for the whole embryo population.
The mass and the initial velocity of the newly formed body are calculated assuming 
perfectly inelastic collision using the center of mass approximation. 

In what follows, we describe briefly how the dissipative forces for type I 
migration have been implemented in our N-body code knowing the surface density 
and temperature profiles of the disk. 

In isothermal disks the total torque $\Gamma$ experienced by a body can be
written as
\begin{equation}
\Gamma/\Gamma_0 = -0.85 - \alpha - 0.9\beta,
\end{equation}
where $\alpha$ and $\beta$ are the negative of the local surface density ($\Sigma(r)$) 
and temperature ($T(r)$) gradients
\begin{equation}
\alpha = -\frac{d\log \Sigma}{d\log r} = -\frac{r}{\Sigma}\frac{d\Sigma}{d r},\ 
\beta =  -\frac{d\log T}{d\log r} = -\frac{r}{T}\frac{d T}{d r},
\end{equation}
and $\Gamma_0 = (q/h)^2\Sigma(r) r^4 \Omega(r)^2$ \citep[see][]{paardekooperetal10}. 
Here $q=m/M_*$ is the body to star mass ratio, $h=H(r)/r$ is the disk constant aspect 
ratio ($H(r)$ being the disk's vertical thickness), and $\Omega$ the Keplerian 
angular velocity. 

Using Equations (1) and (2), the total torque $\Gamma$ can be easily determined,
which enables the calculation of the body's radial migration speed as follows:
\begin{equation}
\dot r = \dot L \left(\frac{d L}{d r}\right)^{-1} = 2r\frac{\Gamma}{L},
\end{equation}
where $L=m\sqrt{GM_*r}$ is the angular momentum of the body. The migration 
timescale is $\tau_\mathrm{migr}=-r/\dot r$, which using Equation (3) reads
\begin{equation} 
\tau_\mathrm{migr}=-\frac{r}{\dot r} =
\frac{1}{m}\frac{h^2M_*^{3/2}}{(1.7+2\alpha+1.8\beta)\Sigma\sqrt{rG}}.
\end{equation} 
Note that a negative $\tau_\mathrm{migr}$ means outward migration.  In
addition to the inward migration, a body also feels the strong damping
effects of the disk on its orbital eccentricity and, since the N-body
part of our model is fully 3-D, its orbital inclination. The
corresponding damping timescales are $\tau_\mathrm{ecc},
\tau_\mathrm{inc}\sim h^4\tau_\mathrm{migr}$ (see the exact formulae in
\citet{tanaka&ward04}).

Knowing the timescales $\tau_\mathrm{migr}$, $\tau_\mathrm{ecc}$, and
$\tau_\mathrm{inc}$, the corresponding forces can be implemented in the
N-body code. In our code we applied the formula of
\citet{cresswell&nelson08}, which for the $i$th body is
\begin{equation}
\mathbf{\ddot r}_{i} = \mathbf{\ddot r}_{i,\mathrm{grav}}
 - \frac{\mathbf{\dot r}_i}{2\tau_\mathrm{migr}}
- 2\frac{(\mathbf{\dot r}_i\cdot\mathbf{r}_i)\mathbf{r}_i}{r_i^2\tau_\mathrm{ecc}}
- \frac{(\mathbf{\dot r}_{i}\cdot\mathbf{k})\mathbf{k}}{\tau_\mathrm{inc}},
\end{equation}
where $\mathbf{k}$ is the unit vector in the $z-$direction and
$\mathbf{\ddot r}_{i,\mathrm{grav}}$ is the gravitational acceleration 
induced by all the other bodies (the N-body force).

\section{Physical background of our simulations}

\begin{figure}[!t]
\plotone{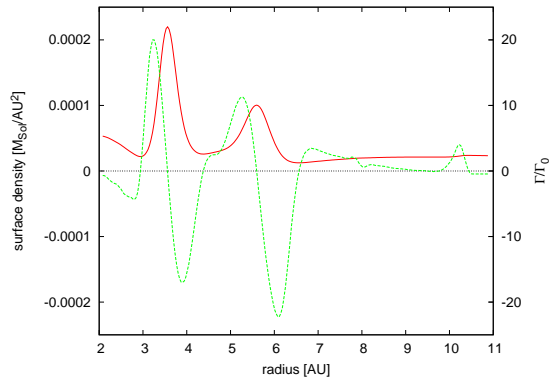}
\caption{The surface density profile plotted with solid red (dark) line, the 
dimensionless torque $\Gamma/\Gamma_0$ with the dotted green 
(light) line. The zero torque places are practically at the two 
``peaks", and the three ``valleys". If $\Gamma/\Gamma_0 > 0$ outward, if 
$\Gamma/\Gamma_0 < 0$ inward migration occur. \label{fig1}}
\end{figure}

As initial conditions of our N-body experiments we used the embryo
population formed during the hydrodynamical simulations of
\citet{lyraetal08,lyraetal09}. In that hydro simulation the initial surface
density and temperature profiles followed power law distributions
$\Sigma\sim r^\alpha$ with $\alpha=-0.5$, and $T\sim r^\beta$ with $\beta =
-1$.  The density profile changed considerably during the simulations, but
due to the local isothermal disk assumption, $\beta=-1$ was kept fixed. For
the torque calculations we used the azimuthally averaged surface density
profile obtained at the end of the hydro simulations, see Fig. 1. To 
keep the problem tractable we fix the gas density profile in time as it
appears in the figure.  The dimensionless torque $\Gamma/\Gamma_0$ is
also displayed. Notice that it reaches values as high as $\pm$10. We
caution that Equation (1) was derived for smooth disks, and may thus not
be unproblematic when used for strongly irregular disks like the one in
our model. But at present this is the only tractable way to treat the
problem of type I migration of many bodies over a long timescale.

In Fig. 1 the two density peaks appear at the inner and the outer edge of
the dead zone, and correspond to the large vortices. Since $p=\Sigma c_s^2$
($p$ is the verticaly integrated pressure and $c_s$ is the local sound
speed), and the temperature distribution ($T(r)$) is fixed, the density
peaks correspond approximately to pressure maxima.

The embryos form in narrow regions around the two pressure peaks
(corresponding to the density peaks of Fig. 1), and they are expected to
disperse somewhat due to mutual gravitational scattering effects.  The peaks
are very nearby to zero-torque radii. They act as planet traps ({\em
``planetary convergence zones''}\footnote{The nomenclature {\it
convergence zone} was suggested by C. Mordasini, and adopted here.})
because inward migration occurs for radii larger than the peak's location,
whereas outward migration occurs for smaller radii than that. There are zero
torque places in the density valleys as well. But, in contrary to the peaks,
they have repulsive character.

\begin{figure*}[!t]
\plotone{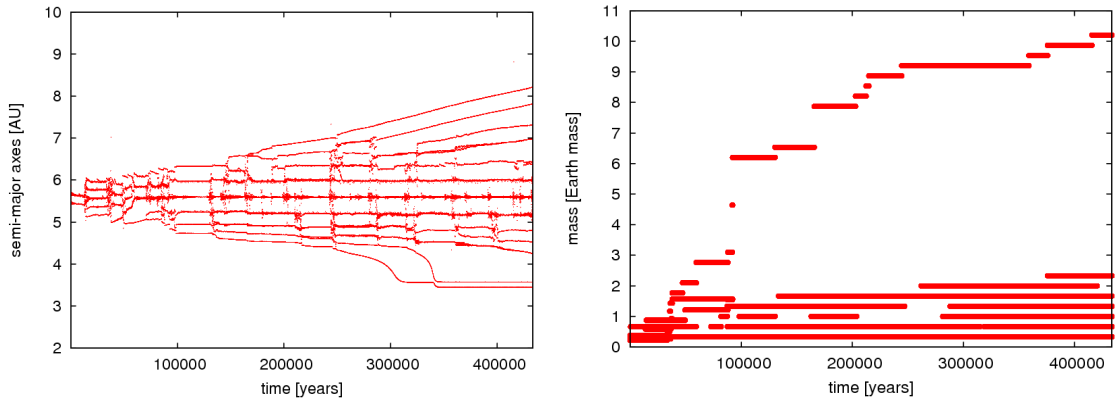}
\caption{\emph{Left}: Evolution of the semi-major axes during the simulation A.
Right: The corresponding overall mass accretion of the embryo population. 
The most massive body is a \emph{planetary core} with mass 
$\approx$ 10 \mearth. \label{fig2}}
\end{figure*}

\begin{figure*}[]
\plotone{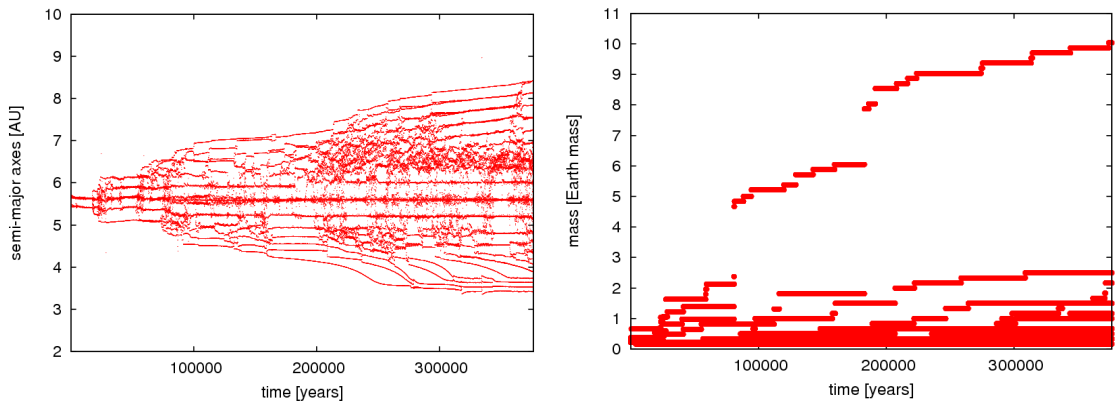}
\caption{The same as in Fig. \ref{fig2} for simulation B.
\label{fig3}}
\end{figure*}

On the other hand, the embryos form on a relatively short timescale, during
200 periods (measured at Jupiter's orbital distance) consuming practically
the solid content of the (strongly truncated) disk used in the hydro
simulations.  If we consider a realistic 100-200 AU disk, repeated formation
of the planetary embryos can be expected as long as the large vortices at
the dead zone edges exist, and the disk contains enough solid material with
size between 1 cm - 1 m. These particles are very strongly affected by the
gas drag, and therefore are drifted inward quickly and continuously to the
embryo forming vortices as long as gas exists in the system. In this way the
outer part of the disk acts as a large reservoir of solid material and due
to the strong radial drift the large vortices, which are working as ``planet
factories" are replenished continuously with ``raw" material. The continuous
embryo formation in the ``planet factory'' is an important part of our
physical assumptions. In our N-body runs we do not intend to model
this process, for feasibility reasons. 
But to mimic the continuous production
of new bodies, we simply inject Mars-mass embryos stochastically into the
pressure trap region, given a production rate which we take as a parameter
of our model. The new bodies are assumed to form only at the outermost
pressure trap (the one at the outer edge of the dead zone). We do this for
two reasons: One is to make the proces more ``clean'' and understandable,
focusing on a single pressure trap only. The other is that it is reasonable
to assume that the influx of dust from the very outer disk regions gets
captured by the outer pressure trap, thus choking the inner pressure trap
off from the supply of new material.

Armed with the above ideas, we performed two long term N-body simulations 
to see how terrestrial planets, and even planetary cores can be formed around
the zero torque places of the disk.

\section{Results and discussion}

Since we are interested in the formation of massive planets, we
considered as initial embryo population of our N-body runs only the 10 most
massive bodies, formed during one of the hydrodynamical simulations of
\citet{lyraetal09} in the outermost pressure trap.  We performed two
simulations: In the first one (simulation A) a 1/3 \mearthp
body\footnote{Corresponding to the mass of the largest body obtained by
\citet{lyraetal09}.} was assumed to form at irregular time intervals
according to a Poisson distribution with a rate of $2\times 10^{-4}$
year$^{-1}$. In the second one (simulation B), a 1/6 \mearthp body
was assumed to form at time intervals also following Poisson distribution
with a rate $4\times 10^{-4}$ year$^{-1}$. The results of simulations A
and B are shown respectively in Figs.~\ref{fig2} and \ref{fig3}.

At the beginning of our N-body simulations close approaches, scattering
events and collisions happened between the initial embryos, after which a
multiple mean motion resonant structure formed. Subsequently, when new
bodies got inserted into the pressure traps, this resonant structure is
perturbed and new scattering events and mergers happen, leading to ever
larger masses. The whole structure of the resonantly interacting planets is
broadening, because of the ever increasing masses of the planets and the
increasing number of them. At around $10^5$ to $1.7\times 10^5$ years the 
planets which have been resonantly pushed the farthest away from the pressure 
bump reach the edges of the type I migration convergence zone 
($4.4 \lesssim r \lesssim 6.6$ AU) and start to migrate {\em away} from their 
birthplace. For the ones that migrate inward, they get {\em again} trapped, 
this time in the innermost migration convergence zone. 

Through collisions some of the bodies are able to increase their masses by
accreting either the newly formed, or the already existing embryos. After
$4\times 10^5$ years-long numerical integration, a planetary core of 10
\mearthp was formed both in simulation A and B. Thus, regarding the final
mass of the planetary core formed, there is no significant difference
between the results of simulations A and B. Comparing the evolution of the
semi-major axes in simulations A and B, it can be clearly seen that in case
A the evolution of the system is ``smoother" than in case B. The reason for
this is that in simulation A a smaller number of bodies is involved than in
simulation B. However, the final mass of the planetary core formed in both
simulations is almost the same after $\sim 4\times 10^5$ years. This means
that the most important parameter of our simulations is the amount of mass
injected during a given time.

We call attention to the fact that, in some aspects, the hydro 
simulations of \citet{morbietal08} are similar to
ours. We therefore briefly compare the methods and results of the two works.
In that study the 2D hydro code FARGO \citep{masset00} has been used, and the N-body
integrator was implemented in 3D, in which the planet's inclination has been
damped according to \citet{tanaka&ward04}. During the hydro simulations the
steep surface density profile halted the migration of 10 embryos (having
each a mass of 1 \mearth), but the resonant structure between them
survived only temporarily, and some of the embryos were able to collide and
merge without any additional perturbative event. In our case, if there would
have been no continuous formation of embryos, the resonant structure would
be stable, preventing the bodies from further collisions.  There are however
a few major differences between the physical models. The static surface
density profile used in our study (providing a strong constant torque),
differs from that in the hydro simulations of \citet{morbietal08}.  They use
an \emph{evolving} 2D disk, in which the back reaction of the bodies to the
disk influences the local gas density distribution, thus the torques are
probably non-constant. The torques are also weaker, since the surface
density distribution is apparently much smoother than in our case. If
instead of a strong static torque, we would use a weaker and oscillating
one, the protecting resonant structure might be destroyed, enabling even a
more rapid formation of a planetary core\footnote{The torque oscillation
  would be due to the fact that the trapping zone in fact consists of a
  couple of large vortices arranged in a circle \citep[see][]{lyraetal08},
  which means that, depending on the azimuthal location of the planet
  compared to the locations of those vortices, the planet may experience
  different torques.}. Compared to \citet{morbietal08} the novelty of our
approach is that we consider the continuous formation of embryos assembled
at the pressure trap, and we study their long-term evolution. We also use
smaller building blocks, and follow the core formation process through a
broader mass range of bodies involved in it.

In the core accretion scenario, the formation of gas giant planets can only
occur if planetary cores are grown still in the gas rich phase. As our
simulations show, the time required for the assembly of a 10 \mearthp
core is $\lesssim 0.5\times 10^6$ years in the framework of our settings. 
Since the lifetime of the gas disk is expected to be $\approx$ 5 Myr, there is
significant time left for accretion of nebular gas, completing the formation
of a gas giant planet \citep{alibertetal05}.

\section{Summary and open questions}

In this Letter we investigated the possibility to form large bodies in
protoplanetary disks at such places where the torques responsible for type I
migration vanish. We assumed that at the edges of the dead zone of the disk
large vortices develop that can collect the (from centimeter to meter sized)
solid content of the disk. Through the self gravity of these overdense
regions of solids, relatively large embryos are formed, with masses up to 
1/3 \mearthp \citep{lyraetal08,lyraetal09}.

Due to the particular surface density distribution obtained from the above
mentioned hydro simulations the embryos stay trapped close to their
birthplaces because the location of the type I migration convergence point
is very close to the location of the pressure trap. The embryos capture each
other into mean motion resonances, entering into a very robust protective
configuration against further collisions. This resonant structure is,
however, perturbed from time to time when a new massive embryo forms in one
of the giant vortices. During these perturbative events, the embryos can
collide and, by merging events, form planetary cores as massive as 10
\mearth. The whole process takes $\lesssim$ 0.5 Myr.

We stress that the proposed scenario should also work in disk models where
the pressure trap emerges by other mechanics, such as at the iceline, as in
\citet{kretke&lin07}; or in zonal flows, as in \citet{johansenetal09}.

Besides cores of giant planets, other massive terrestrial planets can also
be formed, which by scattering at later epochs may lead to the formation of
a complete planetary system. The escape of the bodies from the zero
torque places can also be caused by the slowly forming giant planet, which
by opening a gap will change the surface density profile. 
On the flip side, however, this scenario would
predict that, if the trap is located beyond the ice line, all rocky planets of
the resulting planetary system should be ice- or ocean planets, which 
for our solar system is clearly not the case. Further research is thus
required to investigate these issues.

It is clear, however, that our present model is still very simplified, and
further study is required to verify our proposed scenario for 
the formation of rocky/icy planetary cores. For instance 
\citet{morbietal08} have shown that, due to the
large vortices at the planet trap, the semi-major axes of the embryos
oscillate. This means that the position of the zero torque point of any
given planet is not constant in time. Thus the dynamics of the whole embryo
population is perturbed, which may result in more effective collisions of
bodies at the convergence zone, shortening the time to form a planetary
core. On the other hand, the perturbations induced in the disk by these
planets may shake up and destroy the tranquil nature of the pressure bumps
in which embryos are supposed to be formed, thus perhaps quenching the
formation of new embryos. Also, the planets that are pushed away from the
convergence zone to larger radial distances may form a barrier to dust
drifting inward from the outer parts of the disk. All of these questions
require further study, and much more detailed modeling. But considering that
the combined dust- and planet-migration trap scenario we propose here has
the potential to solve both the meter-size barrier problem and the time
scale problem of oligarchic growth, it seems worthwhile to invest
substantial effort in studying planet formation along these lines. 

We thank the anonymous referee for useful suggestions.

\vspace{0mm}

\end{document}